\documentstyle[12pt]{article}
\begin{document}
\title{ Experimental Constraints on the Neutrino Oscillations and a Simple
Model of Three Flavour Mixing
\thanks{Work supported in part by the KBN-Grants 2-0273-91-01
and 2-0224-91-01.}}
\author{Piotr A. R\c aczka, Andrzej Szymacha\\
Institute of Theoretical Physics, Warsaw Uniwersity,\\
Ho\.{z}a 69, 00-681 Warsaw, Poland. \\ \\
\and
Stanis\l aw  Tatur \\
N.Copernicus Astronomical Center,\\ Polish Academy of Sciences,\\
Bartycka 18, 00-716 Warsaw, Poland.}
\date{}
\maketitle
\begin{abstract}
\noindent A simple model of the neutrino mixing is considered, which
contains only one right-handed neutrino field, coupled via the
mass term to the three usual left-handed fields. This is a
simplest model that allows for three-flavour neutrino
oscillations. The existing experimental limits on the neutrino
oscillations are used to obtain constraints on the two free
mixing parameters of the model. A specific sum rule relating the
oscillation probabilities of different flavours is derived.
\end{abstract}
\newpage
In the past fifteen years a considerable effort has been
made to detect the effect of neutrino oscillations \cite{1}. A positive signal
in
the neutrino oscillation experiments would indicate nonzero neutrino masses
and provide some information on the pattern of the lepton mixing.
Constraints have been obtained on almost all possible types of neutrino
oscillations. The results of the neutrino oscillation experiments are
usually expressed in the form of limits on the mixing angle as a function
of the difference in the squared neutrino masses, under the assumption that
the oscillating transition occurs between only two weak interaction
eigenstates. In this note we consider a model of neutrino oscillations
that goes beyond the two-flavour approximation, allowing for oscillations
involving all three neutrino flavours, but which is much simpler than the
most general case with a three-flavour mixing. This model contains only
one right-handed neutrino field, coupled via the mass term to the three
usual left-handed fields. We show that the formulae for the neutrino
oscillation probabilities in this model may be expressed in a compact form.
We consider constraints on the neutrino mixing implied by the experimental
limits on various oscillation probabilities. We show that these constraints
have a simple geometric interpretation. We discuss in some detail the
constraints from the presently available data from the accelerator
neutrino oscillation experiments. Finally, we obtain in the considered model
a sum rule relating the oscillation probabilities.

The effect of the neutrino oscillation in the vacuum is
described by the formula for the probability $P_{ij}(L)$ of detecting
the weak interaction eigenstate $\nu _{j}$ at the distance $L$
from the region in which neutrinos in the weak interaction
eigenstate $\nu _{i}$ are produced \cite{2} :
\begin{equation}
P_{ij}=\sum_{k}\mid U_{jk}\mid^{2}\mid U_{ik}\mid^{2}+Re\sum_{k\neq l}U_{jk}
U_{jl}^{\star}U_{ik}^{\star}U_{il}\exp[-i\frac{(m_{k}^{2}-m_{l}^{2})L}{2p}]
\end{equation}
where $U_{kl}$ is a unitary mixing matrix relating the electroweak
neutrino eigenstates $\nu _{i}$ with the mass eigenstates
$N_{k}$
\begin{equation}
\nu_{i}=\sum_{k}U_{ik}N_{k}
\end{equation}
and $p$ denotes the neutrino momentum. In (1) it is assumed that the
neutrinos are ultrarelativistic.

If only two neutrino flavors are taken into account, the mixing matrix takes
the form:
\begin{equation}
\left(
\begin{array}{c}
\nu_{1}\\
\nu_{2}
\end{array}
\right)=
\left(
\begin{array}{cc}
\cos\theta&\sin\theta\\
-\sin\theta&\cos\theta
\end{array}
\right)
\left(
\begin{array}{c}
N_{1}\\
N_{2}
\end{array}
\right)
\end{equation}
and the formula (1) reduces to the well known expression \cite{3} :
\begin{equation}
P_{12}(L)=\sin^{2}2\theta_{12}\sin^{2}\frac{\Delta m^{2}L}{4p}
\end{equation}
where $\Delta m^{2}=\mid m^{2}_{1}-m^{2}_{2}\mid $. This formula is
used in most analyses of the experimental data on the neutrino
oscillations \cite{1}. The experimentally determined limits on the
transition probabilities $P_{ij}$ are translated into constraints
on the allowed values of $\sin^{2}2\theta _{ij}$, depending on
the value of $\Delta m^{2}$. As an illustration of the present
experimental situation we show in Fig.1 a compilation of the
best constraints obtained from the accelerator experiments on
the $\nu _{e}-\nu _{\mu }$, $\nu _{e}-\nu _{\tau }$, $\nu _{\mu
}-\nu _{\tau }$ oscillations and the inclusive $\nu _{e}-\nu
_{x}$, $\nu _{\mu }-\nu _{x}$ transitions, where $\nu _{x}$ denotes
neutrino of arbitrary type.
It should be noted however, that it
is impossible for nontrivial $\nu _{e}-\nu _{\mu }$, $\nu _{e}-\nu
_{\tau }$, and $\nu _{\mu }-\nu _{\tau }$ oscillations to have
simultaneously the two-state character. For example, if we
assume that the $\nu _{e}-\nu _{\mu }$ oscillations have a
two-state character, then at the same time we a priori exclude
the possibility of any $\nu _{e}-\nu _{\tau }$, $\nu _{\mu }-\nu
_{\tau }$ oscillations, while the probabilities for the $\nu
_{e}-\nu _{x}$ and $\nu _{\mu }-\nu _{x}$ transitions are
trivially reduced to the $\nu _{e}-\nu _{\mu }$ case.  Therefore
the set of constraints on $\sin^{2}2\theta _{ij}$ given by the
conventional analysis does not reflect properly the patterns of
the neutrino mixing which are still allowed by the available
experimental data.

On the other hand, a general analysis involving three-flavour neutrino
mixing appears to be rather complicated. The three-flavour mixing matrix
contains four free parameters, which may be chosen for example in
the form:
\begin{eqnarray}
U&=&\left(
\begin{array}{ccc}
\cos\varphi&-\sin\varphi&0\\
\sin\varphi&\cos\varphi&0\\
0&0&1
\end{array}
\right)
\left(
\begin{array}{ccc}
\cos\vartheta&0&\sin\vartheta\\
0&1&0\\
-\sin\vartheta&0&\cos\vartheta
\end{array}
\right) \nonumber \\
&  &\hspace{3cm}
\left(
\begin{array}{ccc}
1&0&0\\
0&e^{i\delta}&0\\
0&0&1
\end{array}
\right)
\left(
\begin{array}{ccc}
\cos\psi&\sin\psi&0\\
-\sin\psi&\cos\psi&0\\
0&0&1
\end{array}
\right)
\end{eqnarray}
$\varphi ,\vartheta $ and $\psi $ denote mixing angles and
$\delta $ denotes the CP-violating phase. (This parametrization
is identical to the well known Kobayashi-Maskawa parametrization
\cite{4} except for the permutation of axes and redefinition of $\delta$.)
The experimental constraints on these four parameters depend in general on
two neutrino mass squared differences, which are unknown. Another
complication comes from the fact, that experimentally measured
probabilities are related to the theoretical probabilities via a
convolution with phenomenological functions carrying the
information on the energy spectrum of the neutrino beam or the
geometry of the experimental apparatus. In a general
three-flavour analysis such convolutions would have to be
numerically reevaluated, taking into account three oscillating
terms instead of one as in (4). A three-flavour analysis of
the accelerator and reactor data was attempted in \cite{5}. However,
the results of \cite{5} depend in an essential way on the data from
the Bugey reactor experiment \cite{6}, which seemed to restrict one of
the mass squared differences to a narrow range, thus simplifying
the whole analysis. Unfortunately, the Bugey reactor data was
later found to be inconsistent with several other reactor
experiments \cite{7,8}. A three-flavour analysis may be also found in
\cite{9}, in which the early oscillation experiments are discussed.

It may well be however, that allowing for arbitrary
configuration of neutrino masses introduces an unnecessary complication.
It seems reasonable to expect, that one of the neutrino mass eigenstates
would be much heavier than the remaining two. Such a hierarchical pattern
is observed with other leptons and in the quark sector, and there is an
argument involving the so called see-saw mechanism \cite{10} that neutrino
masses should also have this structure. If this would be the case,
then in the oscillation experiments sensitive to the heavier mass eigenstate
the two other states would appear as effectively massless \cite{9}. This
shows that it would be useful to consider a model of neutrino mixing and
masses in which only one mass eigenstate has mass different from zero.

The neutrino mass term of such a model may be written in the form:
\begin{equation}
L_{\nu-mass}=m(c_{e}\bar{\nu}^{L}_{e}+c_{\mu}\bar{\nu}^{L}_{\mu}+c_{\tau}
\bar{\nu}^{L}_{\tau})N^{R}_{3}+h.c.
\end{equation}
where
\begin{equation}
\mid c_{e}\mid^{2}+\mid c_{\mu}\mid^{2}+\mid c_{\tau}\mid^{2}=1
\end{equation}
The parameters $c_{i}$ in this mass term may be always redefined to
be real (and positive), so that there is no CP-violation. These
parameters may be interpreted as directional cosines which fix
the orientation of the massive eigenstate in the
three-dimensional space spanned by the weak eigenstates:
\begin{equation}
c_{i}=\cos\alpha_{i}
\end{equation}
It should be noted, that the neutrino mass term (6) is a simplest
extension of the Standard Model of electroweak interactions
which allows for nontrivial three-flavour neutrino
oscillations, and therefore it may be of some interest in
itself.

The expressions for the neutrino oscillation probabilities
in this model may be derived by a simple argument. The form of the inclusive
probabilities is immediately obtained when we note, that the neutrino state
vector that was initially $a \nu _{i}$ weak
eigenstate evolves in a two-dimensional space spanned by the
massive component and this weak eigenstate. Therefore:
\begin{equation}
P_{ix}=\sin^{2}2\alpha_{i}\sin^{2}\frac{m^{2}L}{4p}
\end{equation}
The expressions for the exclusive probabilities may then be
determined as a solution of a system of equations:
\begin{eqnarray}
P_{ex}=P_{e\mu}+P_{e\tau}\\
P_{\mu x}=P_{\mu e}+P_{\mu\tau}\\
P_{\tau x}=P_{\tau e}+P_{\tau\mu}
\end{eqnarray}
where we have $P_{ij}=P_{ji}$ because there is no CP-violation in our
model. In this way we find:
\begin{equation}
P_{ij}=4\cos^{2}\alpha_{i}\cos^{2}\alpha_{j}\sin^{2}\frac{m^{2}L}{4p}
\end{equation}
We see that despite the fact, that all types of neutrino oscillations
may be consistently accommodated in our model, the formulae for
the oscillation probabilities have a simple appearance.

The expressions (9,11) may be also obtained from the general
formula (1) using the mixing matrix (5), provided that we identify the
angles $\varphi $ and $\vartheta $ as the two angles fixing the orientation
of the massive state relative to the weak eigenstates, i.e.
\begin{equation}
\cos\alpha_{e}=\cos\varphi\sin\vartheta
\hspace{0.5cm}
\cos\alpha_{\mu}=\sin\varphi\sin\vartheta
\hspace{0.5cm}
\cos\alpha_{\tau}=\cos\vartheta
\end{equation}
The value of the mixing angle $\psi $ may be arbitrary because it
corresponds to a redefinition of the massless states, which does
not affect the oscillation probabilities.

Given the formulae for the oscillation probabilities
we may use now the available experimental data to exclude some values
of the mixing angles $\varphi $ and $\vartheta $. The constraints
obtained from the various experiments acquire a simple geometrical
interpretation when one represents different values of $\varphi $
and $\vartheta $ as points on a unit sphere, corresponding to the
locations of a "tip" of the massive eigenstate vector in the space
spanned by the three weak eigenstates. It is enough to consider $\varphi $
and $\vartheta $ in the range $0^{\circ }-90^{\circ }$. The boundaries of
the relevant triangular region on the sphere $(\varphi =0^{\circ },
 90^{\circ }$ and arbitrary $\vartheta , \vartheta =90^{\circ }$ and
arbitrary $\varphi )$ correspond to the neutrino mixing which has a
purely two-flavour character. From the formula (9) we see, that for
a given mass $m$ the limits on the probabilities of
inclusive $\nu _{i}-\nu _{x}$ transitions exclude regions of the unit
sphere bounded by circles of constant $\sin^{2}2\alpha _{i}$. The
limits on probabilities for $\nu _{i}-\nu _{j}$ oscillations exclude
regions bounded by the line on which the
product $\cos^{2}\alpha _{i}\cos^{2}\alpha _{j}$ is constant - which are
projections of hyperbolas on the sphere - and the side of the spherical
triangle joining the $\nu _{i}$and $\nu _{j}$ corners.

It is important that
the actual numerical value of the constraints on $\sin^{2}2\alpha _{i}$ may
be obtained directly from the limits on $\sin^{2}2\theta _{ex}$ and
$\sin^{2}2\theta _{\mu x}$ extracted in the two-state analysis of the
inclusive $\nu _{e}-\nu _{x}$ and $\nu _{\mu }-\nu _{x}$ oscillation
experiments, provided that the $\Delta m^{2}$ of the conventional
analysis is now understood as the mass squared of the massive eigenstate.
Similarly, the constraints on $4\cos^{2}\alpha _{i}\cos^{2}\alpha _{j}$ are
obtained directly from the limits on $\sin^{2}2\theta _{e\mu },
\sin^{2}2\theta
_{e\tau }$ and $\sin^{2}2\theta _{\mu \tau }$ considered in the conventional
approach to the exclusive $\nu _{i}-\nu _{j}$ oscillation experiments.

As an illustration we show in Fig.2 the pattern of constraints for
$m^{2}=1\;eV^{2}$ that corresponds to the $90\%$ CL limits obtained in the
two-state analysis (see Fig.1). We find that the orientations of the massive
eigenstate, allowed by all the available constraints at this mass, are
confined to three rather small regions of rectangular
shape, lying near the corners of
the spherical triangle. The angular dimensions of these rectangles are
determined directly by the angles $\theta _{ij}$ obtained in the two-state
analysis of the experimental data. For example the size of the region near the
$\nu _{\tau }$ corner is determined by $\theta _{\mu x}$ and $\theta
_{ex}$. ( For a general value of mass the relevant bounds for this region
are given by $min(\theta _{\mu x},\theta _{\mu \tau })$ and $min(\theta
_{ex},\theta _{e\tau })$.) In Fig.2 it may be seen that the limit on
$\nu _{e}-\nu _{\mu }$ oscillations restricts also the strength of the
eventual $\nu _{e}-\nu _{\tau }$ and $\nu _{\mu }-\nu _{\tau }$ oscillations.
This is a manifestation of a correlation between oscillations of different
flavours which is present in the considered model.

In Fig.2 we see, that the various constraints represented on a surface of
the sphere have a simple and highly symmetric form. However, for a detailed
analysis of the experimental results it is more convenient to use a
two-dimensional plot in which $\vartheta $ plays the role of a radial variable
and $\varphi $ remains an angular variable. Such plots have been used in
Fig.3 to show how the available constraints evolve when $m^{2}$ is varied
from $0.1\; eV^{2}$ to $1000\;eV^{2}$. The indicated curves reflect the
$90\%$ CL limits obtained in accelerator experiments [11--15]. (An
exception is the plot for $m^{2}=0.1\;eV^{2}$,
in which the constraint from the Goesgen reactor
experiment \cite{7} on $\bar{\nu }_{e}-\bar{\nu }_{x}$ oscillations has been
included.) The plot for $m^{2}=1000\;eV^{2}$
represents the asymptotic form of the constraints for large neutrino mass. We
see that for $m^{2}\ge 1\;eV^{2}$ the parameters consistent with all the
constraints remain located in the approximately rectangular regions in the
corners of the triangle. However, the size of these regions varies
significantly. In Fig.3 we clearly see changes in the character of the
strongest constraints that determine the size of the allowed regions. For
example in the case of the $\nu _{\tau }$ corner, which seems to be the most
interesting from the phenomenological point of view, the dominant constraints
at $m^{2}=1\;eV^{2}$ come from the inclusive experiments, for $m^{2}=5\;
eV^{2}$ from
$\nu _{\mu }-\nu _{x}$ and $\nu _{e}-\nu _{\tau }$ experiments, and for
$m^{2}\ge
10\;eV^{2}$ from $\nu _{e}-\nu _{x}$ and $\nu _{\mu }-\nu _{\tau
}$ experiments.

It should be noted that the allowed regions of the mixing parameters shown
in Fig.2 and Fig.3 indicate parameters consistent with all constraints
reflecting the $90\%$ CL limits on $\sin^{2}\theta _{ij}$, provided that
these constraints are treated independently. A more precise statistical
analysis of the data within our model would require a consideration of joint
probability distributions for $\varphi $ and $\vartheta $ implied by all the
experimental results. Such an analysis goes beyond the scope of the present
paper.

It is interesting to note, that within the considered model one may obtain
a nontrivial sum rule relating the exclusive oscillation probabilities,
which reflects the fact that there are only two independent mixing parameters
in this case. Indeed, let us denote:
\begin{equation}
P_{ij}=R_{ij}\sin^{2}\frac{m^2L}{4p}
\end{equation}
Then we have:
\begin{equation}
R_{e\mu}R_{e\tau}R_{\mu\tau}(\frac{1}{R_{e\mu}}+\frac{1}{R_{e\tau}}+\frac{1}
{R_{\mu\tau}})^{2}=4
\end{equation}
If a positive signal
for nontrivial $\nu _{e}-\nu _{\mu }$, $\nu _{e}-\nu _{\tau }$ and $\nu _{\mu
}-\nu _{\tau }$ oscillations is obtained, then this sum rule may be used as a
test on the character of the neutrino mixing. If for some $m^{2}$ the
$R^{\exp}_{ij}$ factors, obtained from the experimentally determined
probabilities, would satisfy the sum rule (14) with a good accuracy, then this
would be a strong argument in favor of the presence of a dominant massive
neutrino eigenstate with this mass.

Summarizing we may say, that we have discussed a simplest
model of neutrino mixing which allows for three-flavour neutrino oscillations.
This model is of physical interest because of the expected hierarchical
pattern of the neutrino masses. We have shown that using the experimental
limits on the oscillation probabilities it is easy to obtain constraints on
the two parameters that characterize the neutrino mixing in this model. We
have obtained the domain of the mixing parameters consistent with the
available data on the accelerator neutrino oscillation experiments. We have
found that this domain may be estimated directly from the properly
reinterpreted results of the conventional two-flavour analysis. We have
pointed out that there
exists a sum rule relating the exclusive neutrino oscillation probabilities in
this model, which may be used as a test for the presence of a dominant massive
neutrino state.

\newpage
\begin{center}
Figure captions\\[0.5cm]
\end{center}

\noindent Figure 1:  Experimental limits on the neutrino oscillations
obtained in the accelerator experiments,
expressed in terms of constraints on the mixing
angle (two-state analysis): $\nu _{e}-\nu _{\mu }$ \cite{11},
$\nu _{e}-\nu _{\tau }$ \cite{12,13}, $\nu _{\mu }-\nu _{\tau }$ \cite{12},
 $\nu _{e}-\nu _{x}$ \cite{14}, $\nu _{\mu }-\nu _{x}$ \cite{15}.
All curves correspond to $90\%$ CL limits. The limit on $\bar{\nu }_{\mu
}-\bar{\nu }_{e}$ oscillations \cite{16} has been indicated
when it is stronger than the $\nu _{e}-\nu _{\mu }$ limit.
Also the constraint on the $\bar{\nu}_{e}-\bar{\nu }_{x}$ transitions
from the Goesgen reactor experiment \cite{7} has been
included for completeness. \\

\noindent Figure 2:  Experimental constraints on the orientation of the
massive neutrino state with $m^{2}=1\;eV^{2}$.
The indicated curves correspond to
the $90\%$ CL limits on $\sin^{2}2\theta _{e\mu }$, $\sin^{2}2\theta _{\mu
\tau }$, $\sin^{2}2\theta _{ex}$ and $\sin^{2}2\theta _{\mu x}$ obtained
in the two-state analysis. Dashed regions indicate orientations of the massive
eigenstate consistent with all the indicated constraints.\\

\noindent Figure 3: Constraints on the mixing angles $\varphi $ and
$\vartheta $ for six values of the neutrino mass. The radial variable on
these plots is $\vartheta$,and the angular variable is $\varphi $. The
indicated curves reflect the $90\%$ CL limits on $\sin^{2}2\theta _{ij}$
obtained in the two-state analysis of the accelerator data. Thick lines
surround the regions of the allowed values of $\varphi $ and $\vartheta $
for which the massive state is mostly the $\nu _{\tau }$ weak eigenstate.
\end{document}